\documentclass[a4paper,onecolumn,11pt]{quantumarticle}
\pdfoutput=1
\usepackage[utf8]{inputenc}
\usepackage[english]{babel}
\usepackage[T1]{fontenc}
\usepackage{amsmath}
\usepackage{amsfonts}
\usepackage{amsthm}
\usepackage{amssymb}
\usepackage{hyperref}
\usepackage{physics}
\usepackage{tikz}
\usepackage{lipsum}
\usepackage{float}

\newcommand\blfootnote[1]{%
  \begingroup
  \renewcommand\thefootnote{}\footnote{#1}%
  \addtocounter{footnote}{-1}%
  \endgroup
}

\begin{document}

\title{Efficient charge-preserving excited state preparation with variational quantum algorithms}

\author{Zohim Chandani}
\orcid{0009-0000-9834-8328}
\email{zchandani@nvidia.com}
\affiliation{NVIDIA, Quantum Algorithm Engineering, London, UK}

\author{Kazuki Ikeda}
\orcid{0000-0003-3821-2669}
\email{kazuki.ikeda@umb.edu}
\affiliation{Department of Physics, University of Massachusetts Boston, Boston, MA 02125, USA}
\affiliation{Center for Nuclear Theory, Department of Physics and Astronomy, Stony Brook University, Stony Brook, New York 11794-3800, USA}
\affiliation{Co-design Center for Quantum Advantage, Department of Physics and Astronomy, Stony Brook University, Stony Brook, New York 11794-3800, USA}

\author{Zhong-Bo Kang}
\orcid{0000-0003-3161-0381}
\email{zkang@physics.ucla.edu}
\affiliation{Department of Physics and Astronomy, University of California,
Los Angeles, CA 90095, USA}
\affiliation{Mani L. Bhaumik Institute for Theoretical Physics, University of California,
Los Angeles, CA 90095, USA}
\affiliation{Center for Quantum Science and Engineering, University of California, Los Angeles, CA 90095, USA}

\author{\mbox{Dmitri E. Kharzeev}}
\orcid{0000-0002-3811-6952}
\email{dmitri.kharzeev@stonybrook.edu}
\affiliation{Center for Nuclear Theory, Department of Physics and Astronomy, Stony Brook University, Stony Brook, New York 11794-3800, USA}
\affiliation{Co-design Center for Quantum Advantage, Department of Physics and Astronomy, Stony Brook University, Stony Brook, New York 11794-3800, USA}
\affiliation{Energy and Photon Sciences Directorate, Condensed Matter and Materials Science Division, Brookhaven National Laboratory, Upton, New York 11973-5000, USA}

\author{Alexander McCaskey}
\orcid{0000-0002-0745-3294}
\email{amccaskey@nvidia.com}
\affiliation{NVIDIA, Quantum Computing Architecture, Santa Clara, California, USA}

\author{Andrea Palermo}
\orcid{0000-0001-7218-9568}
\email{andrea.palermo@stonybrook.edu}
\affiliation{Center for Nuclear Theory, Department of Physics and Astronomy, Stony Brook University, Stony Brook, New York 11794-3800, USA}

\author{C.R. Ramakrishnan}
\email{cram@cs.stonybrook.edu}
\affiliation{Department of Computer Science, Stony Brook University, Stony Brook, New York 11794-3800, USA}

\author{Pooja Rao}
\email{porao@nvidia.com}
\affiliation{NVIDIA, Quantum Algorithm Engineering, Santa Clara, California, USA}

\author{Ranjani G. Sundaram}
\email{rasundaram@cs.stonybrook.edu}
\affiliation{Department of Computer Science, Stony Brook University, Stony Brook, New York 11794-3800, USA}

\author{Kwangmin Yu}
\orcid{0000-0003-0826-074X}
\email{kyu@bnl.gov}
\affiliation{Computational Science Initiative, Brookhaven National Laboratory, Upton, New York 11973-5000, USA}
\blfootnote{All authors are ordered alphabetically.}
\vspace{-2cm}
\begin{abstract}
Determining the spectrum and wave functions of excited states of a system is crucial in quantum physics and chemistry. Low-depth quantum algorithms, such as the Variational Quantum Eigensolver (VQE) and its variants, can be used to determine the ground-state energy. However, current approaches to computing excited states require numerous controlled unitaries, making the application of the original Variational Quantum Deflation (VQD) algorithm to problems in chemistry or physics suboptimal. In this study, we introduce a charge-preserving VQD (CPVQD) algorithm, designed to incorporate symmetry and the corresponding conserved charge into the VQD framework. This results in dimension reduction, significantly enhancing the efficiency of excited-state computations. We present benchmark results with GPU-accelerated simulations using systems up to 24 qubits, showcasing applications in high-energy physics, nuclear physics, and quantum chemistry. This work is performed on NERSC's Perlmutter system using NVIDIA's open-source platform for accelerated quantum supercomputing - CUDA-Q. 

\end{abstract}
\maketitle
\section{Introduction}

High-performance computing (HPC) can tackle a wide range of scientific challenges, including large-scale simulations, data analysis, and advanced mathematical computations. Powerful computing infrastructure is essential for tasks that require immense processing power and substantial memory resources.

In contrast, quantum computers can address several problems that are challenging for classical computers. Exhaustive searches, optimization, or the simulation of quantum many-body systems are domains where quantum computers can outperform classical computing paradigms~\cite{feynman2018simulating}. 

Considering the complementary strengths of HPC and quantum computing, their integration is essential to leverage their strengths and maximize their potential ~\cite{ALEXEEV2024666,Robledo-Moreno:2024pzz}. By developing hybrid computing infrastructures, it is possible to exploit the capabilities of both systems, expanding their applicability to a wider range of tasks. This integration is especially important in the current Noisy Intermediate-Scale Quantum (NISQ) era~\cite{Preskill2018quantumcomputingin}, where quantum computers are still in the developmental stage and not completely error-free. During this transitional phase, HPC systems can be used to simulate and validate quantum algorithms before they are executed on quantum hardware. This preliminary step enables us to avoid the effects of high error rates and limited coherence times whilst experimenting with algorithmic techniques unimpeded. 

The VQE algorithm, a hybrid quantum-classical approach, involves both quantum and classical computations~\cite{Peruzzo:2013bzg,kandala2017hardware,lanyon2010towards,mccaskey2019quantum,cerezo2021variational,PhysRevA.102.062612,2024arXiv240109253N}. The quantum portion prepares and measures the resulting quantum state, while the classical portion optimizes the parameters of the quantum circuit to guide the algorithm down the cost landscape to minimize the energy. Efficient and precise desired state preparation is crucial for the success of quantum computation, as it directly impacts the algorithm’s convergence and the quality of the results. While VQE stands out as one of the most thoroughly studied algorithms for simulating ground states in the near term, it faces significant hurdles. The primary challenges include difficulties in achieving convergence during optimization, also known as barren plateaus~\cite{Larocca:2024plh,Wang:2020yjh}, and accurately measuring the target energy because of noise and decoherence. These issues can lead to prolonged execution time and the risk of getting trapped in local minima, thereby hindering the algorithm’s efficiency and effectiveness. HPC systems can significantly enhance the classical optimization phase of VQE by efficiently managing large parameter spaces. This integration of QPU + GPU yields faster convergence in the results, ultimately improving the overall performance of the VQE algorithm.

In addition to finding the ground state energy, VQE can be extended to compute excited energy states which was introduced as the variational quantum deflation (VQD) algorithm~\cite{Higgott2019variationalquantum,PhysRevResearch.4.013173}. Excited states of quantum many-body systems are essential for determining their spectra, which provide critical insights into the properties and behaviours of materials and molecules, such as their electronic, vibrational, and rotational states. Accurately computing these excited states is crucial for applications in quantum chemistry, condensed matter physics, and nuclear physics. 


In this work, we introduce the charge-preserving VQD (CPVQD) algorithm which computes excited states in a specific symmetry sector associated with the conserved charge. This addresses some of the limitations of existing methods thereby enhancing the reliability and precision of the results. 
We demonstrate the application of CPVQD to solve problems in quantum chemistry and nuclear physics, such as spectra and excited states of the system.


We utilize CUDA-Q~\cite{cudaq}, NVIDIA’s high-performance platform for accelerating hybrid quantum-classical computing, which enables the development of algorithms for quantum and classical disparate architectures within a single environment. CUDA-Q is well suited for this work since it allows a single program to access CPUs, GPUs, and QPUs, and also ensures that code can be scaled from running on single chips to GPU supercomputers with negligible modification.

\section{\label{sec:CPVQD}Charge-Preserving VQD}

Let $H$ be a Hamiltonian of a many-body system of our interest. We can obtain the excited states of $H$ by VQD whose cost function is given as follows~\cite{Higgott2019variationalquantum}: 
\begin{equation}
\label{eq:cost}
F(\lambda_k)=\bra{\psi(\lambda_k)}H\ket{\psi(\lambda_k)}+\sum_{i=0}^{k-1}\beta_i|\bra{\psi(\lambda_k)}\psi(\lambda_i)\rangle|^2,
\end{equation}
where $\beta_i>0$ are real values, $\lambda_k$ represent the variational parameters, with $k$ indexing the eigenstates of the Hamiltonian. Specifically $k=0$ denotes the ground state while $k>0$ corresponds to the $k$-th excited state. When the variational algorithm converges successfully, the state $\ket{\psi(\lambda_k)}$ approaches to the $k$-th excited state of the quantum many-body system.  As evident from the definition, the cost function of the VQE can be derived by setting $\beta_0=0$. This adjustment simplifies the optimization process, allowing the algorithm to focus on minimizing the energy of the ground state while effectively isolating the contributions of higher excited states.


In principle, we can obtain all energy eigenstates and spectra by using the cost function in equation ~\eqref{eq:cost}, however, this process necessitates meticulous consideration of the charge sectors to ensure that only the relevant physical states are included in the analysis. In fact, not all eigenstates of the Hamiltonian correspond to physical states, even if the Hamiltonian itself is the correct description of the phenomena at hand. This discrepancy arises because the Hilbert space used for qubit representation includes states with different charges. For a faithful simulation, it is essential to exclude spurious states and focus only on the correct charge sector, thus ensuring spectra to be physically meaningful. This selection can be achieved by projecting the Hamiltonian onto the appropriate charge sector, isolating the desired eigenstates and spectra. This procedure also improves the efficiency of the quantum computation.

We propose two different methods to execute CPVQD efficiently, as illustrated below:  
\begin{enumerate}
    \item[Method 1:] We aim to reduce the Hilbert space of the Hamiltonian through the transformation $H \to \widetilde{H}$, where $\dim(H) \geq \dim(\widetilde{H})$ and $\widetilde{H}$ represents the reduced Hamiltonian that retains the essential properties of interest. This is done by fist projecting the Hamiltonian to $H_P=PHP$, and then the dimensions of the matrix are reduced by eliminating the rows and columns of $H_P$ that have been projected out, thus obtaining $\tilde{H}$. By performing the CPVQD to $\tilde{H}$, we can simplify the computational complexity by reducing the size of the problem, while also focusing only on the relevant states. This approach is particularly efficient to solve quantum chemistry and physics problems when we want to perform targeted analysis of quantum systems in a fixed symmetry sector, such as a fixed charge sector, of both ionized and charge neutral molecules.  We will take this approach for our applications to quantum chemistry (Sec.~\ref{sec:chem}) and quantum physics (Sec.~\ref{sec:phys}). There the Hamiltonian $H_P$ corresponds to the eq.~\eqref{eq:block_hamiltonian}, where it is block-diagonalized using the charge operator. The number of qubits we can reduce from the original system is evaluated by eq.~\eqref{eq:reduced_dim} and shown in Fig.~\ref{fig:reduced_charge_sector} for different charges $q$. 
    
    \item[Method 2:] In contrast to dimensional reduction, we can also access a specific symmetry sector by adding constraints to the Hamiltonian. This is achieved through the transformation $H \to \widetilde{H} = H + \alpha H'$, where $\alpha$ is a non-negative number and $H'$ is a Hamiltonian that incorporates the given constraints. See eq.~\eqref{eq:m1} for example. This technique is commonly used to solve optimization problems where the constraints are generally very complex. Unlike the dimensional reduction method, this approach does not alter the number of qubits involved. Although the additional constraint may aggravate the convergence of the algorithm, this method would be useful for maintaining the integrity of the original system while ensuring that the constraints are effectively applied.
\end{enumerate}

The CPVQD algorithm can be applied to a variety of systems across chemistry, biology, physics, and optimization problems. This is particularly valuable in scenarios where the excited states play a crucial role, such as in photo-chemical reactions, non-equilibrium dynamics, particle collisions, thermal systems, and statistical mechanics. In this article, we showcase the applications of the CPVQD specifically in the fields of quantum chemistry and nuclear physics. Although we focus on the natural extension of the original VQD, our method can be applied to various extensions of similar methods, such as SSVQE~\cite{PhysRevResearch.1.033062} and ADAPT-VQE~\cite{Feniou:2023gvo,Farrell:2023fgd}.

\section{Experimental Setup}

This work has been performed using CUDA-Q~\cite{cudaq}, NVIDIA’s open-source high-performance programming platform. CUDA-Q is a flexible, user-friendly programming environment, providing access to CPU, GPU and QPU hardware, alongside performant quantum simulation capabilities. These factors, alongside its wide support of qubit modalities, makes it perfectly suited for work developing novel algorithms and applications. For these experiments, we utilized CUDA-Q’s GPU-accelerated state vector simulator (nvidia-fp64 backend) without any noise modelling. The simulations were executed using a single NVIDIA A100 GPU on the National Energy Research Scientific Computing Center’s (NERSC) Perlmutter supercomputer.

\begin{figure}[h!]
    \centering
    \includegraphics[width=0.45\linewidth]{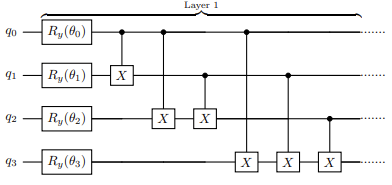}
    \caption{A single layer of the four-qubit ansatz.}
    \label{fig:ansatz}
\end{figure}
\if{
\begin{figure}[h!]
    \centering
    \includegraphics[width=\linewidth]{ansatz.png}
    \caption{The ansatz circuit for four qubits.}
    \label{fig:ansatz}
\end{figure}
}\fi

A circuit diagram for a single layer of the ansatz is depicted in Fig. \ref{fig:ansatz}. It consists of a layer of $R_y$ gates followed by a maximal entangling layer with each qubit being entangled to each other. After applying $L$ layers of the ansatz, a layer of $R_y$ gates is added to all qubits. Therefore, the number of parameters of ansatz is $(L + 1) \times N$ when $L$ and $N$ are the number of layers and the number of qubits, respectively.
Since we used four layers, the number of parameters (optimization space dimension) is  $5N$.

In our experiments, we adopt the Constrained Optimization by Linear Approximation (COBYLA) optimizer, a numerical method for constrained problems where the objective function's derivative is unknown \cite{powell1994direct}.
The COBYLA algorithm has mainly three parameters: the initial change of the variables (rhobeg), the convergence accuracy, and the maximum iteration number. The initial change of the variables (rhobeg) determines the size of the initial ``trust region'' around the starting point. The convergence accuracy determines the condition to stop the iterations. The maximum iteration specifies the maximum number of iterations the algorithm can perform before terminating. Since the convergence accuracy and the maximum iteration determine the stop condition of the iteration, if one condition is satisfied, then the iteration stops. In general, smaller values of the initial change of the variables will lead to more accurate results, but may also increase the computational cost of the algorithm.
In our experiment, we use $0.5$ for the initial change of the variables (an internal option COBYLA) and $625 \times N$ maximum iteration numbers when we have $N$ qubits.
We do not set the convergence accuracy in order to observe the optimization values up to the maximum iteration number.


We implemented the algorithms in Qiskit and CUDA-Q and conducted cross-validation for cases of less than 10 qubits. After that, we ran CUDA-Q for larger system sizes. The workflow in CPVQD is similar to other variational methods where the expectation value from the quantum circuits is used to compute the cost function and the circuit parameters are then optimized via a classical optimizer. In our setup, the cost function evaluation in CUDA-Q \cite{cudaq} has been interfaced with the COBYLA optimizer implemented in SciPy \cite{scipy}.

\section{\label{sec:chem}Application to Quantum Chemistry}
\subsection{\label{sec:sppedup}Dimensional reduction method to compute the full spectra of a fixed charge sector}
As a concrete example of our method, here we consider the applications to quantum chemistry. We use the second quantized form of the electronic Hamiltonian given as follows~\cite{helgaker2013molecular,szabo1996modern}: 
\begin{equation}
\label{eq:Ham_chemistry}
    H=\sum_{ij}h_{ij}a^\dagger_i a_j+\frac{1}{2}\sum_{ijkl}h_{ijkl}a^\dagger_i a^\dagger_ja_ka_l,
\end{equation}
where $a^\dagger_i$ and $a_i$ are the fermionic creation and annihilation operators, $h_{ij}$ and $h_{ijkl}$ are one-electron and two-electron integrals. Those coefficients are determined by the basis set we work with which in our case is STO-3G.

We use the charge operator, $Q$, to select the charge sector:
\begin{equation}
\label{eq:charge_op}
    Q=\sum_i a^\dagger_ia_i.
\end{equation}
Using the Jordan-Wigner transformation~\cite{Jordan:1928wi}, the charge operator --the $z$-component of spin-- is generally written as $Q=\frac{1}{2}\sum_iZ_i$, where $Z_i$ is the Pauli $Z$ operator defined at the $i$-th site. Consequently, the charge is characterized by the disparity between the number of 0's and 1's in a given quantum state, and it takes values between $-\frac{N}{2}$ and $\frac{N}{2}$, where $N$ is the number of qubits associated to the molecule. 

To obtain the eigenstates with charge $q$, one may perform the conventional VQD with respect to the following Hamiltonian:
\begin{equation}
\label{eq:m1}
    \widetilde{H}=H+\alpha (qI-Q)^2. 
\end{equation}
This represents a natural choice, achievable with minimal modifications to the existing setup, and aligns with Method 2 as described in Section~\ref{sec:CPVQD}. Although this is the most intuitive approach, our goal is to find the more efficient solution by exploiting the system's inherent symmetry to decrease the dimensionality of the Hilbert space. It is also important to note that $\alpha$ is a positive value determined heuristically. Consequently, this heuristic determination could increase the likelihood that the VQD algorithm does not converge.

Now we illustrate the dimensional reduction method.  Since the original Hamiltonian~\eqref{eq:Ham_chemistry} conserves the total electric charge ($[H,Q]=0$), we can diagonalize it in the basis of $Q$ in such a way that 
\begin{equation}
\label{eq:block_hamiltonian}
    H=\begin{pmatrix}
        H_{q=Q_{\max}}&&&\\
        &\ddots&&&\\
        &&H_{q=0}&&&\\
        &&&\ddots&\\
        &&&& H_{q=Q_{\min}},
    \end{pmatrix}
\end{equation}
where $H_q$ is the block-diagonal Hamiltonian in the charge $q$ sector and $Q_{\max}=-Q_{\min}=N/2$. 


In general, the charge $q$ of a state depends on the number $n=0,\dots, N$ of 1s in the state vector as $q=N/2-n$. In what follows, let $\mathcal{H}_q$ be the Hilbert space associated with the charge $q$. The number of basis vectors of a subspace of fixed charge is equal to the number of ways we can choose the positions of the $n$ 1's in our size $N$ vector, that is:
\begin{equation}
\label{eq:reduced_dim}
    \dim\left(\mathcal{H}_{q=\frac{N}{2}-n}\right)={}_{N}C_{n}={}_{N}C_{N/2-q}.
\end{equation}
where ${}_nC_k=\binom{n}{k}=\frac{n!}{(n-k)!k!}$. For instance, charge 0 sector is the largest subspace, with $\dim\left(\mathcal{H}_{q=0}\right)={}_{N}C_{N/2}$, consisting of the states having the same number $(N/2)$ of 1s and 0s. The Neel state $\ket{\text{Neel}}=\ket{0101\cdots01}$ is a typical example of a chargeless state, and other vectors of the basis of this subspace can be created as:
\begin{equation}
    \left\{\prod_{i=0}^n X_{\sigma(i)}\ket{\text{Neel}}:n\in\{0,2,4,\cdots,N\},\sigma\in \mathfrak{S}_\text{even}\times\mathfrak{S}_\text{odd}\right\}, 
\end{equation}
where $\mathfrak{S}_\text{even/odd}$ are the permutation groups of the even/odd sites. On the other hand, the for $q=\pm N/2$ only one state is present, i.e. $|1,1\dots,1\rangle$ or $|0,0\dots,0\rangle$.

\begin{figure}[H]
    \centering
    \includegraphics[width=\linewidth]{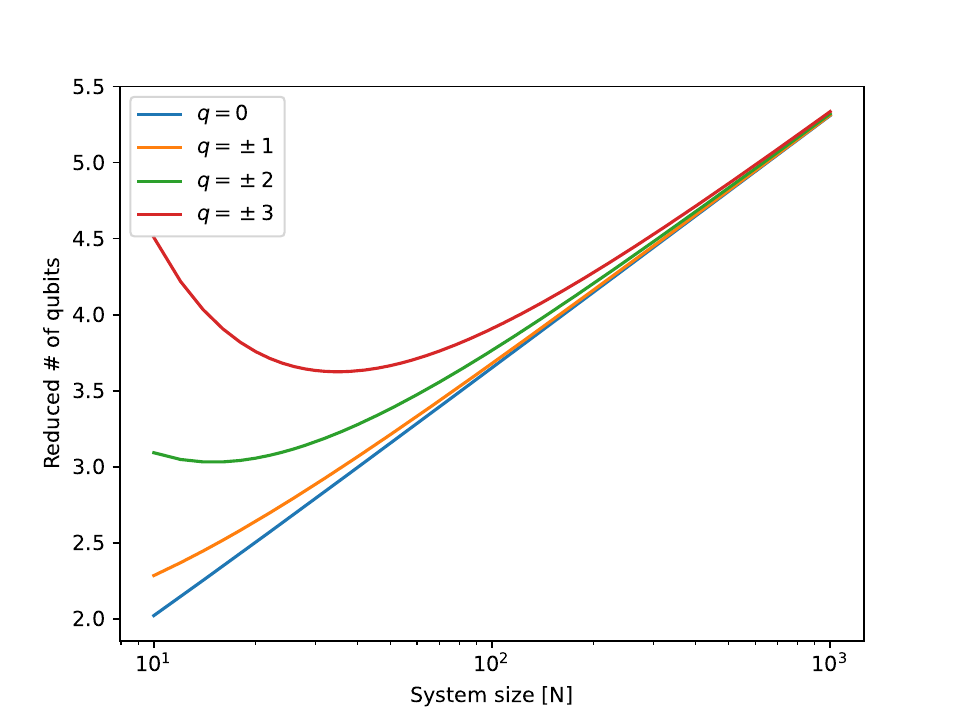}
    \caption{Number of reduced qubits with respect to system size $N$ for various charges $q$. }
    \label{fig:reduced_charge_sector}
\end{figure}

The number of qubits necessary to simulate the charge $q$ sector is the smallest integer $N_*$ larger than or equal to $\log_2 \dim(\mathcal{H}_{q})$, that is 
\begin{equation}
    N_*-1<\log_2 \dim(\mathcal{H}_{q})\le N_*. 
\end{equation}
Therefore the dimensional reduction is useful to simulate ions, which we will address in Sec.~\ref{sec:ion}. In Fig.~\ref{fig:reduced_charge_sector} we show the reduced number of qubits for different charge sectors $q=0,\pm1,\pm2,\pm3$. As a trivial case, the maximal/minimal charge sectors can be represented by only 1 qubit regardless of $N$, by mapping $\ket{0\cdots0}\to\ket{0},\ket{1\cdots1}\to\ket{1}$. 

To define the molecular Hamiltonian~\eqref{eq:Ham_chemistry} of our interest, let us explain the basis of the molecules so that we can implement them on a quantum circuit. One of the simplest types of basis sets used in quantum chemistry are the STO-$n$G basis sets, which stands for Slater Type Orbital-$n$ Gaussians~\cite{10.1063/1.1672392}. In these basis sets, each atomic orbital is approximated by a Slater Type Orbital (STO). The STOs themselves are represented using a combination of $n$ Gaussian Type Orbitals (GTOs). Due to their simplicity, STO-$n$G basis sets are often referred to as minimal basis sets. This is because they include only the orbitals necessary to describe the Hartree–Fock (HF) state and other orbitals of similar energy.

However, calculations performed with minimal basis sets are generally of limited accuracy, providing only a qualitative description of the system under study. It is crucial to understand that when performing a HF calculation using an STO-$n$G basis set, the true HF energy — defined as the energy obtained from a grid-based method on an infinitely precise grid — cannot be achieved. This limitation arises because STO-$n$G basis sets only approximate the true HF orbitals. Despite these limitations, STO-$n$G basis is still a useful benchmark for quantum chemistry study implementable on a near-term quantum device~\cite{RevModPhys.92.015003,2018arXiv181209976C} and there are various quantum platforms such as OpenFermion~\cite{McClean:2017ims}, Qiskit Aqua~\cite{qiskit2024}, and QDK-NWChem~\cite{2019arXiv190401131H} 

\begin{figure}[H]
    \centering
    \includegraphics[width=\linewidth]{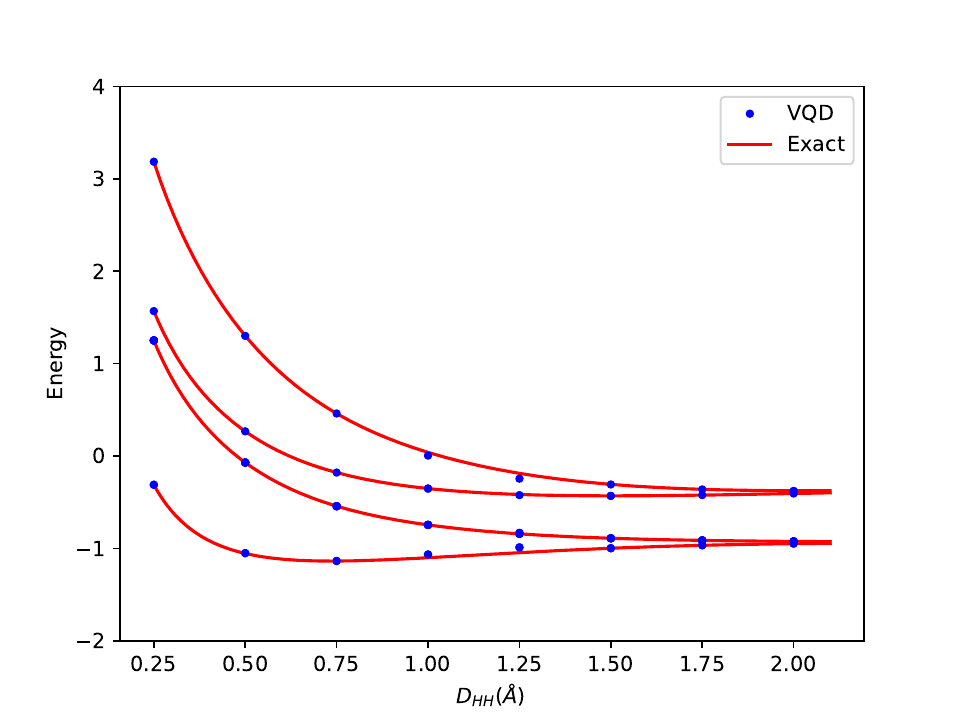}
    \caption{The bond-length dependence of the spectrum (in Hartree) of $H_2$ in the charge 0 sector, where all energy spectra carry charge 0 in the STO-3G basis. The red curves and the blue dots are obtained by exact diagonalization and VQD, respectively.}
    \label{fig:H2}
\end{figure}

In what follows, we take the minimal basis set called STO–3G. In this basis the Hamiltonian of $H_2$ can be represented by $4$ qubits as confirmed by the data from OpenFermion, however, the charge 0 sectors is spanned by these six basis vectors:
\begin{equation}
\label{eq:charge0}
    \ket{1010}, \ket{1001}, \ket{0110},\ket{1100},\ket{0011},\ket{0101}. 
\end{equation}
Therefore 3 qubits are enough to simulate it. The results of $H_2$ in the charge 0 sector is plotted in Fig.~\ref{fig:H2}. 

Moreover, if there is additional symmetry that commutes with both Hamiltonian $H$ and the charge $Q$, we can further reduce the dimension of the Hilbert space. As an example, let us consider the case where the system has the parity symmetry (the reflection with respect to the center of the lattice). Here it is important that the charge and the parity operator commute, therefore the Hamiltonian can be simultaneously diagonalizable on the same basis. 

\begin{figure}[H]
    \centering
    \includegraphics[width=\linewidth]{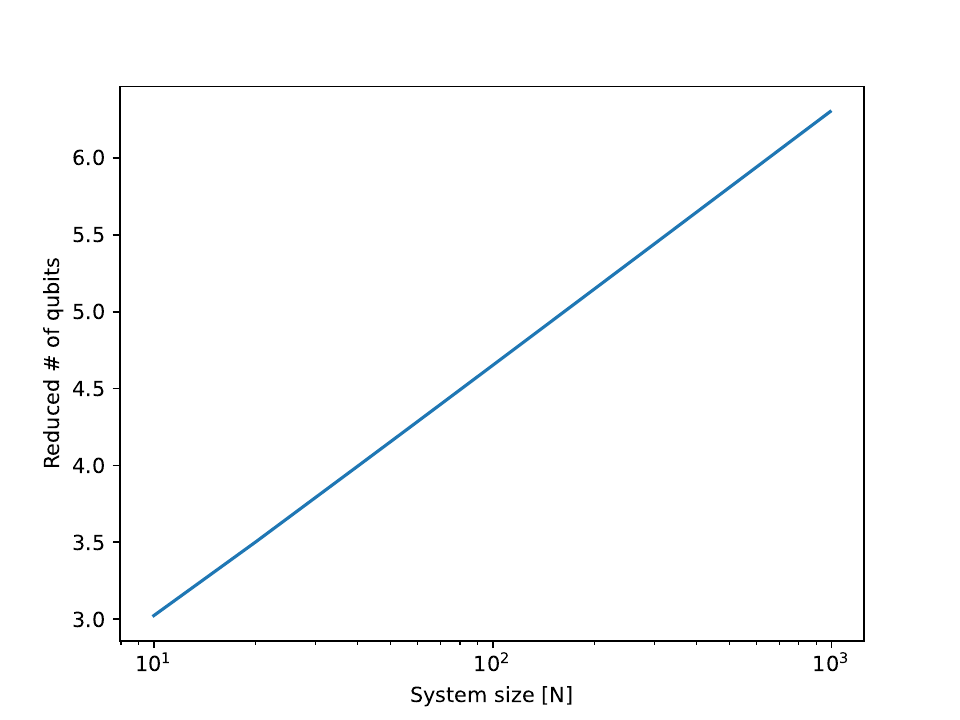}
    \caption{The reduced number of qubits as a function of the system size evaluated according to eq.\eqref{eq:reduced}.}
    \label{fig:reduced_qubits}
\end{figure}

Imposing the additional symmetry is useful to reduce the system further. First of all, the charge 0 sector is always written by an even number of qubits. When $N$ is not multiple of 4, then the number of the chargeless parity symmetric $(q=0, P=+1)$ basis is ${}_{N}C_{N/2}/2$, which counts the symmetric combinations states with the same number of 0s and 1s: e.g. $(|101010\rangle+|010101\rangle)/\sqrt{2}$. When $N$ is multiple of 4, there are also self-symmetric states like $\ket{1001},\ket{0110}$ and the number of such states is ${}_{N/2}C_{N/4}$. Therefore, we can summarize the dimension of the charge 0 and parity even sector as follows: 
\begin{equation}
\label{eq:dim0P1}
    \dim(\mathcal{H}_{q=0,P=1})=
    \begin{cases}
        \frac{1}{2}\left({}_NC_{N/2}+{}_{N/2}C_{N/4}\right)& N\equiv0\mod4\\
        \frac{1}{2}\left({}_NC_{N/2}\right)&N\equiv2\mod4
    \end{cases}
\end{equation}
As a consequence, the dimension of the Hilbert space can be reduced exponentially, therefore the number of qubits required for the simulations is reduced logarithmically: 
\begin{equation}
\label{eq:reduced}
    N-\log_2(\dim(\mathcal{H}_{q=0,P=1})).
\end{equation}
To confirm this we plot the reduced number of qubits in the log-scale of the original system size $N$ (Fig.~\ref{fig:reduced_qubits}). 

According to~\cite{Kühn2019}, the number of two qubit gates (CNOTs) in VQE increases logarighmically with respect to system size $N$, so we can at least reduce the complexity of the conventional VQD logarithmically for a single spectrum and a fixed system size. Moreover, the requisite system size for implementing the Hamiltonian also diminishes logarithmically.

\subsection{\label{sec:ion}Efficient Quantum Simulation of Ions}
Dimensional reduction proves to be highly advantageous for simulating ions, which are charged molecules.  This is because the dimensions of the Hilbert space of the charge $q$ sector decreases as the absolute value of $q$ increases, simplifying the study of the ion’s electronic structure and reactivity. 

To illustrate this concept, let us consider the spectra of the helium hydride ion, $\text{HeH}^+$, which carries a charge of $q=+1$. It is known that the Hamiltonian of $\text{HeH}^+$ can be represented by 4 qubits. At first glance, one might expect this 4-qubit system to yield $2^4 = 16$ possible spectra. However, the $q = +1$ charge sector is spanned only by 4 basis vectors: $\ket{1000},\ket{0100},\ket{0010},\ket{0001}$. Consequently, the dimensionality of the system can be reduced from 16 to 4 in the $q=+1$ sector, and the system can be simulated accurately with only 2 qubits. 

\begin{figure}[H]
    \centering
    \includegraphics[width=\linewidth]{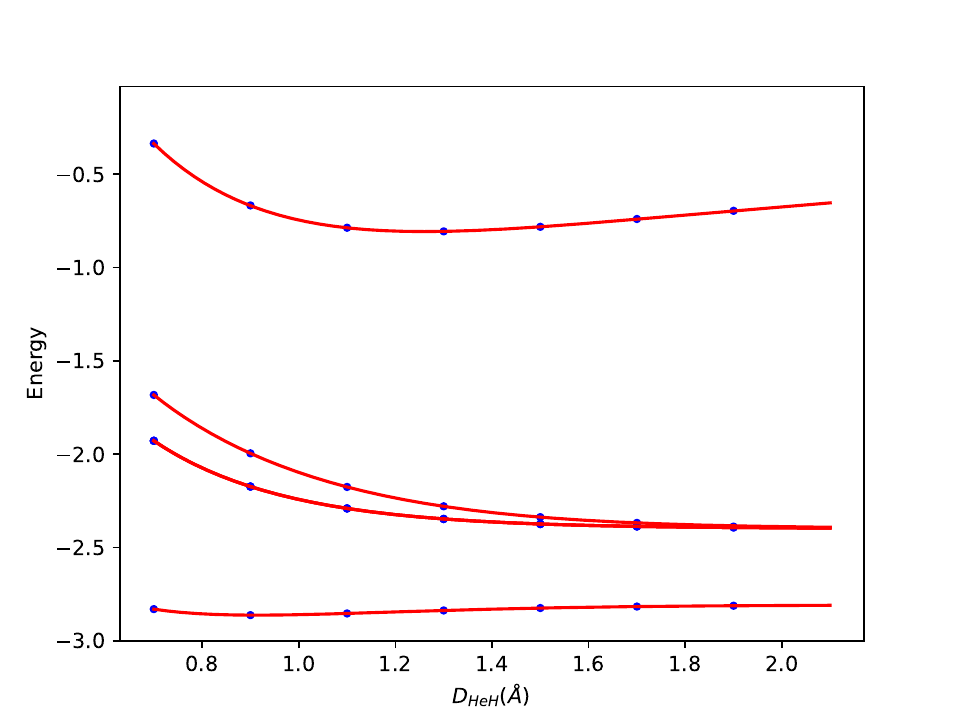}
    \caption{The bond-length dependence of the spectra (in Hartree) of $\text{HeH}^+$ in the charge 1 sector. All charge +1 spectra in the STO-3G basis are shown here. The red curves and the blue dots are obtained by exact diagonalization and CPVQD, respectively.}
    \label{fig:HeH}
\end{figure}

Fig.~\ref{fig:HeH} shows the results of the charge preserving VDQ. Here, we show the the distance dependence of $\text{HeH}^+$ energy spectrum. Again, the solid lines were obtained by exact diagonalization of the Hamiltonian and the dots corresponds the energy eigenvalues gained by the charge-conserving VQD. 

\section{\label{sec:phys}Application to Quantum Physics}
\subsection{Preliminaries}
To demonstrate the wider applicability of our technique, we apply it to simulations of high-energy and nuclear physics phenomena. We consider (1+1)-dimensional quantum field theories (QFT) of massive Dirac fermions in this section. Unlike the quantum chemistry examples discussed in Sec.~\ref{sec:chem}, these models allow us to investigate systems with more degrees of freedom under various conditions, including different coupling strengths and boundary conditions.
Moreover, these theories can be easily extended to more complex systems with nontrivial dynamics, various flavors, and gauge bosons. For example, the $\mathbb{Z}_3$ gauge theory in (1+1)-dimensions serves as an appropriate application of our method for analyzing baryon structure, effectively modeling large $N$ QCD in (1+1) dimensions~\cite{THOOFT1974461,PhysRevD.13.1649,PhysRevLett.95.250601,PhysRevD.110.045013,2024PhRvD.109k4502H}. This approach would be meaningful for deepening our comprehension of higher-dimensional quantum gauge theories, including the (3+1)-dimensional quantum chromodynamics (QCD). Since the simulation of a gauge theory requires a significant amount of qubits, reducing this number is useful to perform quantum simulation more efficiently. 

In Sec.~\ref{sec:massless}, we introduce a basic model of Dirac fermions in (1+1)-dimensions under the PBC, which generates the symmetries including the $\mathbb{Z}_2$ symmetry and the translation symmetry, which is a continuous $U(1)$ symmetry associated with the conservation of momentum. In Sec.~\ref{sec:QED}, we extend our model to quantum electrodynamics (QED), which is a $U(1)$ gauge theory and called Schwinger model. The calculation of excited states in this model is important to obtain the phase diagram of the model, as well as to explore many interesting properties of the gauge theory such as quasi-parton structure~\cite{PhysRevLett.110.262002,Grieninger:2024axp,Grieninger:2024cdl}, for example.

In Sec.~\ref{sec:massgap}, we perform VQD to compute the mass gap in the model for $q=0$. 
Our goal here is to show the agreement between the theory and VQD results. It is important to note that the unphysical 1st excited states are degenerated in an $N$-fold way, which makes it nontrivial to obtain the correct physical excited states even at a small mass. Our VQD  procedure to obtain excited physical states (Fig.~\ref{fig:VQD_m025}) resolves this degeneracy.

\subsection{\label{sec:massless}A massive Dirac model}
Let us start with the most basic case with the minimum parameter. Here, we consider a massive Dirac fermion model in the (1+1) dimensions whose Lagrangian density is given as follows:
\begin{align}
\label{eq:massive_Dirac}
\mathcal{L}=\bar\psi\left(i\gamma^\mu\partial_\mu-m\right)\psi.
\end{align}
This model has is $\mathbb{Z}_2$ symmetry, corresponding to the conservation of the particle number. Using the staggered fermion~\cite{Susskind:1976jm,Kogut:1974ag} and the Jordan-Wigner transformation~\cite{Jordan:1928wi}, the corresponding qubit Hamiltonian with the periodic boundary condition can be written as 
\begin{align}
\begin{aligned}
\label{eq:Ham_pbc}
H=&\frac{1}{4a}\sum_{n=1}^{N-1}(X_{n}X_{n+1}+Y_{n}Y_{n+1})+\frac{m}{2}\sum_{n=1}^{N}(-1)^nZ_n\\
&+\frac{(-1)^{\frac{N}{2}}}{4a}\left(X_{N}X_1+Y_{N}Y_1\right)\prod_{n=2}^{N-1}Z_n, 
\end{aligned}
\end{align}
where $N$ is even. The first term corresponds to the hopping, the second is the mass term, and the last term is the hopping with the periodic boundary condition. Throughout the work, we use the following Dirac matrix notation: $\gamma^0 = Z$, $\gamma^1 = i,Y$, and $\gamma^5=\gamma^0 \gamma^1 = X$. 

The PBC gives the $\mathbb{Z}_2$-symmetry, which is discussed by the operator $P'= \prod_{j} Z_j$~\cite{wei2011global}, counting the total number of particles in the system. To pick up the $P'=+1$ states from $H$, one can perform VQD using the following Hamiltonian as introduced in Method 2: 
\begin{equation}
    \widetilde{H}=H+\alpha(1-P')^2, 
\end{equation}
Here it is important although the model has even number of particles, the Hilbert space contains the $P'=-1$ basis. In principle, one can obtain the correct spectra with $\tilde{H}$, however one can notice that the Hilbert space can be reduced.  Indeed the Hamiltonian can be block diagonalized by the eigenstates of the parity operator:
\begin{equation}
    UHU^\dagger= \begin{pmatrix}
        H_+&O\\
        O& H_-
    \end{pmatrix},
\end{equation}
where $U$ is a unitary matrix, $H_{+},H_-$ are the Hamiltonians which have $\pm1$ eigenvalues of $P'$, respectively. In this case, it is not difficult to create the basis vectors that make the Hamiltonian block diagolaized. 
For instance, it is straightforward to check that the following set of the basis vectors are parity even:  
\begin{equation}
\label{eq:parity_even}
    \left\{\prod_{i=0}^n X_{\sigma(i)}\ket{0\cdots0}:n\in\{0,2,4,\cdots,N\},\sigma\in \mathfrak{S}_N\right\},
\end{equation}
where $\mathfrak{S}_N$ is the set of all permutations of the lattice sites $\{0,1,\cdots,N-1\}$. The dimension of this subspace is $2^{N/2}$.

\subsection{\label{sec:QED}Quantum Electrodynamics (QED)}
Here we put an electric field into the previous model~\eqref{eq:massive_Dirac} and consider the quantum electrodynamics (QED), which is called the Schwinger model~\cite{Schwinger:1962tp} and has been widely used as a benchmark model of quantum simulation of a gauge theory~\cite{Farrell:2023fgd,Klco:2018kyo,Butt:2019uul,Magnifico:2019kyj,Shaw:2020udc,Kharzeev:2020kgc,Ikeda:2020agk,2023arXiv230111991F,PhysRevD.107.L071502,Ikeda:2023zil,Ikeda:2023vfk, Florio:2024aix,Ikeda2024,PhysRevD.109.114510}.  Its Lagrangian density is given as follows:
\begin{equation}
\label{eq:L0}
    \mathcal{L} = -\frac{1}{4}F_{\mu\nu}F^{\mu\nu}+\bar{\psi}(i\gamma^\mu\partial_\mu-g\gamma^\mu A_\mu-m)\psi. 
\end{equation}

Applying the Jordan-Wigner transformation as previously done, we find that the qubit Hamiltonian of the model is
\begin{align}
\label{eq:Ham}
\begin{aligned}
     H=&\frac{1}{8a}\sum_{n=1}^{N}(X_n X_{n+1}+Y_n Y_{n+1})+\frac{m}{2}\sum_{n=1}^N(-1)^n Z_n+\frac{a\ g^2}{2}\sum_{n=1}^{N} L^2_n,
\end{aligned}
\end{align}
where $L_n$ is the electric field operator 
\begin{equation}
    L_n=\sum_{j=1}^{n}\frac{Z_j+(-1)^j}{2}, 
\end{equation}
which satisfies the Gauss law constraint 
\begin{equation}
    L_{n+1}-L_n=aQ_n. 
\end{equation}
where $Q_n$ is the local electric charge density operator at the $n$-th site represented as follows:
\begin{align}
    Q_n \equiv \,& \bar{\psi}\gamma^0\psi_n = \frac{Z_n+(-1)^n}{2a}. 
\end{align}
We define the total electric charge operator $Q \equiv a\sum_{n=1}^{N} Q_n$, which commutes with the Hamiltonian. 

Note that we can reduce the dimensionality of the problem by enforcing the Hamiltonian to be charge neutral, as we did in Fig.~\ref{fig:reduced_qubits}, and parity even. 
Throughout the rest of this work, we will fix the lattice spacing $a$ and the gauge coupling $g$ to $ag=1$. Here both $1/a$ and $g$ have the mass dimension, therefore $ag$ and $m/g$ are dimensionless.  We evaluate energy in the unit of $ag$ in Fig.~\ref{fig:VQD_m025}.

\subsection{\label{sec:massgap}Mass gap in the (1+1)d-QED}
At the large mass limit of the massive Schwinger model, the Hamiltonian is dominated by the mass term 
\begin{equation}
\label{eq:mass_term}
    H_m=\frac{m}{2}\sum_{n=0}^{N-1}(-1)^{n-1}Z_n, 
\end{equation}
whose ground state is the Neel state $\ket{\text{Neel}}=\ket{0101\cdots01}$ and the ground state energy is 
\begin{equation}
    \bra{\text{Neel}}H_m\ket{\text{Neel}}=\frac{-mN}{2}
\end{equation}

In general, the $n$-th excited states can be prepared by 
\begin{equation}
\label{eq:excited}
    \prod_{i=1}^n X_{k_i}\ket{\text{Neel}},
\end{equation}
where $k_i$ is an element of $\{0,\cdots,N-1\}$ such that $k_i\neq k_j$ when $i\neq j$. The corresponding energy eigenvalue is
\begin{equation}
\label{eq:excited_energy}
    E_n=\frac{-m(N-2n)}{2}. 
\end{equation}

For example, the 1st excited states have the $N$-fold degeneracy 
\begin{equation}
    X_k\ket{\text{Neel}}~(k=0,\cdots,N-1)
\end{equation}
and the 1st excited state energy is 
\begin{equation}
   \bra{\text{Neel}}X_kH_mX_k\ket{\text{Neel}}=\frac{-m(N-2)}{2} 
\end{equation}
for all $k=0,\cdots, N-1$. 

However not all eigenstates are physically allowed states, since physical states must respect the Gauss law constraint.  In the Schwinger model, the correct physical states $\ket{\psi}$ are charge neutral:
\begin{equation}
    \bra{\psi}Q\ket{\psi}=0, 
\end{equation}
where $Q=\frac{1}{2}\sum_{n=0}^{N-1}Z_n$ is the charge operator, which is exactly the same operator we used for quantum chemistry~\eqref{eq:charge_op}. Therefore among all excited states given by eq.~\eqref{eq:excited}, a state is physical if and only if $n$ is even, otherwise the states are charged. Hence, the 1st excited states are not physical, but the 2nd excited states are physical. Therefore the physical 1st excited state corresponds to the 2nd excited state of the Hamiltonian. By using eq.~\eqref{eq:excited_energy}, the mass gap at the large mass limit is obtained as
\begin{equation}
\label{eq:massgap}
    (E_2-E_0)/g=2m/g. 
\end{equation}

\begin{figure}[H]
    \centering
    \includegraphics[width=\linewidth]{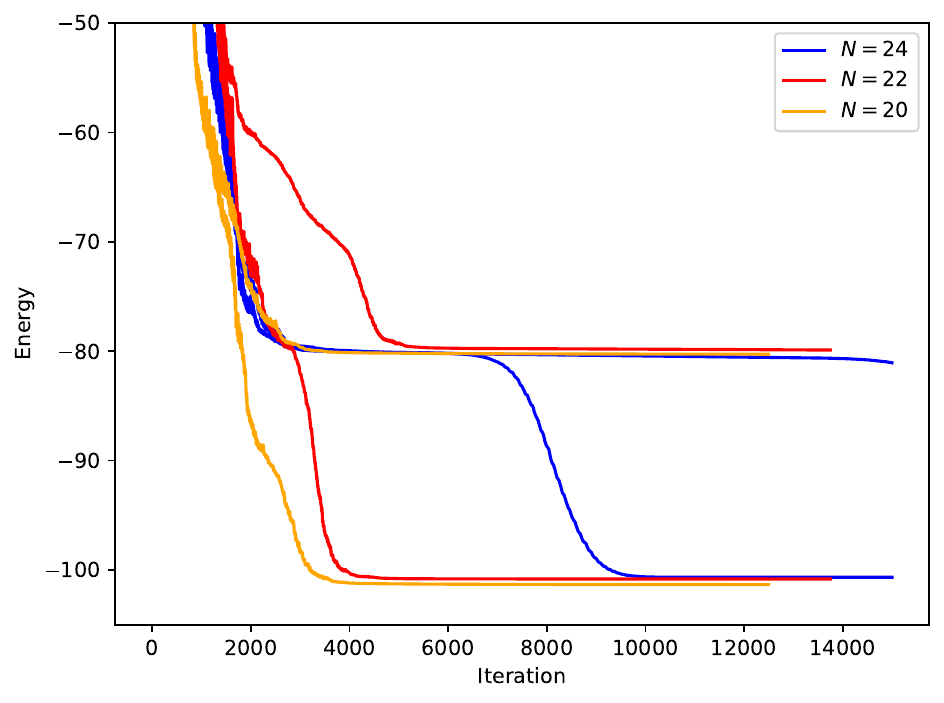}
    \caption{CPVQD results of the Schwinger model in the charge 0 sector, with a large mass $m/g=10$ up to $N=24$. The plots with the same color correspond to the same system size $N$. The gap corresponds to the gap between the ground state energy and the physical $1^{st}$ excited energy. Here unphysical spectra carrying a charge are successfully removed, therefore the correct mass gap $(\approx 2m/g=20)$ is realized, as expected by eq.~\eqref{eq:massgap}. Here the energy is dimensionless as we evaluate it in the unit of $ag=1$, which is dimensionless.}
    \label{fig:VQD_m025}
\end{figure}

On the other hand, in the small mass limit, the squared rest mass is
\begin{equation}
    m_\eta^2=m_S^2+m_\pi^2=\frac {g^2}\pi-\frac{m\langle \overline{\psi}\psi\rangle_0}{f^2}
\end{equation}
with $f=1/\sqrt{4\pi}$ the $\eta$ decay constant~\cite{Grieninger:2023ufa}. 
The vacuum chiral condensate is finite
in the chiral limit, with
$\langle\overline\psi \psi\rangle_0=-\frac{e^{\gamma_E}}{2\pi}\frac{g}{\sqrt{\pi}}$, 
where $\gamma_E=0.577$ is Euler constant~\cite{Sachs:1991en,Steele:1994gf}. In this case, the degeneracy in the 1st excited state revolves, and there are $N$ distinct eigenstates between the ground state and the physical 1st excited states. In other words, the $E_{N+1}-E_0$ corresponds to the physical mass-gap. If we perform the original VQD, we need to calculate those unnecessary $N$ spectra before getting the correct eigenstate. This step can be removed by imposing the charge constraint on the cost function of the VQD. In Fig.~\ref{fig:VQD_m025} we show CPVQD results up to $N=24$ qubits of the Hamiltonian at a large mass. In the figure, the ground state energy and the physical 1st excited state energy carrying the charge 0 are shown. Here we leave a remark on the small mass case: To recover the continuum limit of the theory, the mass term should be modified to $m_{lat}=m-ag^2/8$~\cite{PhysRevResearch.4.043133}. 

We measured the execution time of the quantum circuit simulation (QC) and the classical optimization routine (CC) on Perlmutter of the National Energy Research Scientific Computing Center, a DOE Office of Science User Facility supported by the Office of Science of the U.S. Department of Energy.
Perlmutter is an HPE (Hewlett Packard Enterprise) Cray EX supercomputer based on the HPE Cray Shasta platform.
Perlmutter consists of 1536 GPU-accelerated nodes with 1 AMD Milan processor equipped with 4 NVIDIA A100 GPUs, and 3072 CPU-only nodes with 2 AMD Milan processors. Table \ref{tab:running_time} summarises the results. The circuit simulation and measurement was performed using CUDA-Q with the GPU accelerated statevector simulation target on the A100 GPU. The classical optimization routine was performed on the CPU cores of the node.

In the first column, QC and CC represent the quantum computing part and the classical part, respectively.
In the second row, GS and ES represent the ground state computation and the exited state computation, respectively.
As the number of qubits increases, the ratio of the quantum part increases and the ratio of the classical part decreases.
Since the dimension of the optimization domain increases linearly with respect to the number of qubits, the computing complexity of the optimizer is at most $\mathcal{O}(n^3)$ when $n$ is the number of qubits \cite{powell1994direct, powell1998direct, conn2009introduction}.
On the other hand, the computing complexity of the quantum state evolution increases exponentially in classical computers since the dimension of the quantum state increases exponentially.
Hence, the running time of the quantum part increases exponentially as the number of qubits increases linearly.
The computing complexity analysis explains why the ratio of the quantum parts in Table \ref{tab:running_time} increases as the number of qubits increases.
Also, this shows the potential of quantum speed-up when the quantum circuit is executed on quantum computers instead of classical quantum simulators (CUDA-Q in our experiments).

\begin{table}
\centering 
\renewcommand{\arraystretch}{1.3} 
\begin{tabular}{ c | c | c | c | c | c | c }
 \hline\hline
  N (qubits) & \multicolumn{2}{|c|}{20}   & \multicolumn{2}{|c|}{22} & \multicolumn{2}{|c}{24} \\ 
  \hline
   &  GS & ES & GS & ES & GS & ES \\
  \hline \hline
 QC (sec) & 361.13  & 2646.20 & 1632.72  & 5643.40 & 7011.20 & 17068.56 \\
 \hline
 CC (sec) & 8.90 & 34.71 & 13.23 & 38.89 & 18.42 & 50.06   \\
 \hline
 Total (sec) & 370.03 & 2680.91 & 1645.95  & 5682.30 & 7029.62  & 17118.62  \\
 \hline
 QC/Total (\%) & 97.59  & 98.70 & 99.20  & 99.32  & 99.73 & 99.71 \\
 \hline
 CC/Total (\%) & 2.41 & 1.29 & 0.80 & 0.68  & 2.6 & 0.29 \\
 \hline\hline
\end{tabular}
\caption{The running time of CPVQD. This is the running time of the experiments described in Fig. \ref{fig:VQD_m025}. GS and ES represent the ground state and the excited state computing time, respectively. QC (sec) and CC (sec) represent the quantum part and the classical part running time, respectively, in seconds. The quantum part includes building the parameterized ansatz, executing the quantum circuit (ansatz), and the cost function computation based on the quantum state (or measurement). The  classical part includes optimization of the computed cost function.}
\label{tab:running_time}
\end{table}

\section{Conclusion and Discussion}
The determination of excited states within quantum systems is a cornerstone for advancing numerous fields, including but not limited to, optical spectroscopy, and the dynamics of particle interactions. While advancements in quantum computing, like the Variational Quantum Eigenvalue Solver (VQE), have effectively found ground state energies with reduced circuit depths, calculating excited states has remained a challenge, often necessitating high-depth controlled-unitaries or significantly expanded sampling. In this study, we introduce benchmark results for the GPU-accelerated CPVQD algorithm, extending its application to systems with up to 24 qubits. This marks a significant leap forward in computational capability in domains such as high energy physics, nuclear physics, and quantum chemistry. By applying dimension reduction techniques, specifically exploiting parity and charge conservation, we enhance the efficiency of excited state computation. Our methodology computes the excited state energies by optimizing a modified cost function, operating seamlessly on NERSC’s Perlmutter using NVIDIA’s CUDA-Q. This work not only underscores the potential of CPVQD in handling large, complex quantum systems but also sets the stage for future innovation in the quantum computation landscape, enabling more precise simulations and analyses across scientific disciplines.

In this paper, we have unveiled our inaugural benchmark results. Our methodology can be seamlessly extended to encompass various analogous approaches, such as SSVQE and ADAPT-VQE~\cite{PhysRevResearch.1.033062,Feniou:2023gvo,Farrell:2023fgd}. These extensions promise to offer a novel and significant applications of quantum computers to the analysis of quantum many-body systems, particularly those involving excited state phenomena in both equilibrium and non-equilibrium systems.

\section*{Acknowledgment}
We thank Jin-Sung Kim for useful communications about CUDA-Q. The authors also thank Sebastian Grieninger and Ismail Zahed for useful discussions about the mass gap of the Schwinger model, and Fangcheng He for the productive discussion of the projection operators. This work was supported by the U.S. Department of Energy, Office of Science, National Quantum Information Science Research Centers, Co-design Center for Quantum Advantage (C2QA) under Contract No.DE-SC0012704 (KI, DK), as well as the National Science Foundation under grant No. PHY-1945471 (ZK). This research used resources of the National Energy Research Scientific Computing Center, a DOE Office of Science User Facility supported by the Office of Science of the U.S. Department of Energy under Contract No. DE-AC02-05CH11231 using NERSC award
DDR-ERCAP0028999.
This research used quantum computing resources of the Oak Ridge Leadership Computing Facility, which is a DOE Office of Science User Facility supported under Contract DE-AC05-00OR22725.

\bibliographystyle{quantum}
\bibliography{main}

\end{document}